\documentclass{PoS}

\title{Field-strength correlators for QCD in a magnetic background}

\ShortTitle{Field-strength correlators for QCD in a magnetic background}

\author{\speaker{Enrico Meggiolaro}\\
   Dipartimento di Fisica, Universit\`a di Pisa, and INFN, Sezione di Pisa,\\
   Largo Pontecorvo 3, I-56127 Pisa, Italy\\
   E-mail: \email{enrico.meggiolaro@unipi.it}}

\author{Massimo D'Elia\\
   Dipartimento di Fisica, Universit\`a di Pisa, and INFN, Sezione di Pisa,\\
   Largo Pontecorvo 3, I-56127 Pisa, Italy\\
   E-mail: \email{massimo.delia@unipi.it}}

\author{Michele Mesiti\\
   Dipartimento di Fisica, Universit\`a di Pisa, and INFN, Sezione di Pisa,\\
   Largo Pontecorvo 3, I-56127 Pisa, Italy\\
   E-mail: \email{mesiti@pi.infn.it}}

\author{Francesco Negro\\
   Dipartimento di Fisica, Universit\`a di Pisa, and INFN, Sezione di Pisa,\\
   Largo Pontecorvo 3, I-56127 Pisa, Italy\\
   E-mail: \email{fnegro@pi.infn.it}}

\abstract{We present the results of an exploratory study (by means of Monte
Carlo simulations on the lattice) of the properties of the gauge-invariant
two-point correlation functions of the gauge-field strengths for $N_f=2$ QCD
at zero temperature and in the presence of a magnetic background field:
the analysis provides evidence for the emergence of anisotropies in the
nonperturbative part of the correlators and for an increase of the gluon
condensate as a function of the external magnetic field.}


\FullConference{38th International Conference on High Energy Physics\\
                3-10 August 2016\\
                Chicago, USA}

\usepackage{amsmath}

\newcommand{\beq}{\begin{eqnarray}}
\newcommand{\eeq}{\end{eqnarray}}

\newcommand{\Tr}{\ensuremath{\mathrm{Tr}}}

\newcommand{\Dmrns}{{\cal D}_{\mu\rho,\nu\sigma}}

\def\spose#1{\hbox to 0pt{#1\hss}}
\def\ltapprox{\mathrel{\spose{\lower 3pt\hbox{$\mathchar"218$}}
 \raise 2.0pt\hbox{$\mathchar"13C$}}}

\begin{document}

\section{Introduction}
\label{intro}

In the last few years, an increasing interest has arisen in the scientific
community about the study of the strong interactions in the presence of
strong magnetic fields (see, e.g.,~Ref.~\cite{lecnotmag}).

From a phenomenological point of view, the physics of some compact
astrophysical objects, like magnetars, 
of noncentral heavy ion collisions 
and of the early Universe 
involve the properties of quarks and gluons in the presence of magnetic
backgrounds going from $10^{10}$ Tesla up to $10^{16}$ Tesla
(i.e., up to $|e| B \sim 1$ GeV$^2$).

From a purely theoretical point of view, one emerging feature
is that gluon fields, even if not directly coupled to electromagnetic 
fields, can be significantly affected by them:
effective QED-QCD interactions, induced by quark loop contributions,
can be important, because of the nonperturbative nature of the 
theory 
(see, e.g., Refs.~\cite{ozaki,Ilgenfritz:2012fw,reg2,Ilgenfritz:2013ara,struzzo}).

It is well known (see Ref. \cite{DDSS02} for a complete review on this subject)
that many nonperturbative properties of the QCD vacuum can be usefully
parametrized 
in terms of the gauge-invariant two-point field-strength correlators, defined as:
\beq
\Dmrns(x) = g^2 \langle
\Tr [ G_{\mu\rho}(0) S(0,x) G_{\nu\sigma}(x) S^\dagger(0,x) ]
\rangle ,
\label{defcorr}
\eeq
where $G_{\mu\rho} = T^aG^a_{\mu\rho}$ is the field-strength tensor
($T^a$ being the $SU(3)$ fundamental generators),
and $S(0,x)$ is the parallel transport from $0$ to $x$ along a straight line,
needed to make the correlators gauge invariant.
Such correlators were first considered to take into account the nonuniform
distributions of the vacuum condensates and their effects on the levels
of the $Q\overline{Q}$ bound states. 

Here we present the results of an exploratory lattice study \cite{DMMN2016}
(performed for $N_f = 2$ QCD at zero temperature)
of the effects of a magnetic background field on these gluon-field correlators.
The analysis is focused on a quantity of phenomenological 
interest which can be extracted from the correlators,
the so-called {\it gluon condensate}.

\section{Field-strength correlators in the absence or presence of external fields}
\label{structure}

In the vacuum and in the absence of external sources, Lorentz symmetry
($SO(4)$ symmetry in the Euclidean space)
implies a simple form for the two-point functions in Eq.~(\ref{defcorr}), 
which can be expressed in terms of two independent scalar functions
of $x^2$, which are usually denoted by ${\cal D} (x^2)$ and
${\cal D}_1 (x^2)$ 
(see Ref. \cite{DDSS02} and references therein):
\beq
\label{usual_param}
\lefteqn{
\Dmrns(x) =
(\delta_{\mu\nu}\delta_{\rho\sigma} - \delta_{\mu\sigma}\delta_{\rho\nu})
\left[ {\cal D}(x^2) + {\cal D}_1(x^2) \right] } \nonumber \\
& & + (x_\mu x_\nu \delta_{\rho\sigma} - x_\mu x_\sigma \delta_{\rho\nu}
+ x_\rho x_\sigma \delta_{\mu\nu} - x_\rho x_\nu \delta_{\mu\sigma})
\frac{\partial{\cal D}_1(x^2)}{\partial x^2} .
\eeq
The presence of an external field breaks Lorentz/$SO(4)$ symmetry,
so that the most general parametrization is more
complex than the one reported in Eq.~\eqref{usual_param}.
A detailed discussion of this problem can be found in Ref. \cite{DMMN2016}
and will not be reported here.

On the other hand, in our investigation on the lattice
we have considered only correlators of the kind
$\mathcal{D}_{\mu\nu,\xi}(d) \equiv \mathcal{D}_{\mu\nu,\mu\nu}(x = d\hat{\xi})$,
where the two plaquettes are parallel to each other and the separation $x$ is
along one ($\hat{\xi}$) of the four basis vectors of the lattice
[$\hat{x}=(1,0,0,0)$, $\hat{y}=(0,1,0,0)$, $\hat{z}=(0,0,1,0)$,
$\hat{t}=(0,0,0,1)$].
These amount, in general, to $24$ different correlation functions. Without any 
additional external field, the symmetries of the system group these $24$
correlators into two equivalence classes,
usually denoted as $\mathcal{D}_\parallel$ (when $\xi=\mu$ or $\xi=\nu$ ) and
$\mathcal{D}_\perp$ (when $\xi\neq\mu$ and $\xi\neq\nu$), with
${\cal D}_\parallel = {\cal D} + {\cal D}_1
+ x^2 \frac{\partial {\cal D}_1}{\partial x^2}$
and ${\cal D}_\perp = {\cal D} + {\cal D}_1$.

In the presence of a constant and uniform magnetic field $\vec{B}$ oriented
along the $z$ axis ($\vec{B} = B \hat{z}$), the $SO(4)$ Euclidean symmetry
breaks into $SO(2)_{xy} \otimes SO(2)_{zt}$.
By virtue of this residual symmetry (which implies two equivalence relations,
one between the two {\it transverse} directions $\hat{x} \sim \hat{y}$
and another between the two {\it longitudinal} [or: ``{\it parallel}'']
directions $\hat{z} \sim \hat{t}$),
the 24 correlation functions $\mathcal{D}_{\mu\nu,\xi}$ are grouped
into $8$ equivalence classes, which can be denoted as:
\beq
\mathcal{D}^{tt,t}_\parallel ,~~~~
\mathcal{D}^{tt,p}_\perp ,~~~~
\mathcal{D}^{tp,t}_\parallel ,~~~~
\mathcal{D}^{tp,p}_\parallel ,~~~~
\mathcal{D}^{tp,t}_\perp ,~~~~
\mathcal{D}^{tp,p}_\perp ,~~~~
\mathcal{D}^{pp,t}_\perp ,~~~~
\mathcal{D}^{pp,p}_\parallel ,
\label{corrclasses}
\eeq
where the superscripts $t$ and $p$ stand respectively for the
\emph{transverse} ($\hat{x},\hat{y}$) directions and for the
``\emph{parallel}'' ($\hat{z},\hat{t}$) directions.


In the absence of external field ($B=0$), the correlators were
directly determined by numerical simulations on the lattice in Refs.
\cite{DP1992,npb97,plb97,DDM2003}, using the following
parametrization vs. the distance $d$:
$\mathcal{D} = A_0 e^{-\mu d} + \frac{a_0}{d^4}$,
$\mathcal{D}_1 = A_1 e^{-\mu d} + \frac{a_1}{d^4}$;
that is, in terms of $\mathcal{D}_\parallel$ and $\mathcal{D}_\perp$:
\beq
\label{oldparam}
\mathcal{D}_\parallel = \left[ A_0+ A_1 \left(1-\frac{1}{2}\mu d\right)
\right] e^{-\mu d} + \frac{a_\parallel}{d^4} ,~~~~
\mathcal{D}_\perp = \left(A_0+ A_1\right) e^{-\mu d} + \frac{a_\perp}{d^4} ,
\eeq
where $a_\parallel = a_0-a_1$ and $a_\perp = a_0+a_1$.
The terms $\sim 1/d^4$ are of perturbative origin and (according to the
{\it Operator Product Expansion}) are necessary to describe the short-distance
behavior of the correlators.
The exponential terms represent, instead, the nonperturbative contributions:
in particular, the coefficients $A_0$ and $A_1$ can be directly linked to the
{\it gluon condensate} of the QCD vacuum (see Eq. \eqref{g2def} below).

Inspired by the parametrization (\ref{oldparam}) used in the case $B=0$,
we have used for the eight functions (\ref{corrclasses}) in the
case $B\neq 0$ the following parametrization:
\beq
\label{newparam}
&&\mathcal{D}^{tt,t}_\parallel =
\left[ A^{tt}_0+ A^{tt}_1 \left(1-\frac{1}{2}\mu^{tt,t}d\right) \right]
e^{-\mu^{tt,t}d} + \frac{a^{tt,t}_\parallel}{d^4} ,~~~~
\mathcal{D}^{tt,p}_\perp =
(A^{tt}_0+ A^{tt}_1) e^{-\mu^{tt,p}d} + \frac{a^{tt,p}_\perp}{d^4} ,\nonumber\\
&&\mathcal{D}^{tp,t}_\parallel =
\left[ A^{tp}_0+ A^{tp}_1 \left(1-\frac{1}{2}\mu^{tp,t}d\right) \right]
e^{-\mu^{tp,t}d} + \frac{a^{tp,t}_\parallel}{d^4} ,\nonumber\\
&&\mathcal{D}^{tp,p}_\parallel =
\left[ \tilde{A}^{tp}_0+ \tilde{A}^{tp}_1 \left(1-\frac{1}{2}\mu^{tp,p}d\right)
\right] e^{-\mu^{tp,p}d} + \frac{a^{tp,p}_\parallel}{d^4} ,\nonumber\\
&&\mathcal{D}^{tp,t}_\perp =
(A^{tp}_0+ A^{tp}_1) e^{-\mu^{tp,t}d} + \frac{a^{tp,t}_\perp}{d^4} ,~~~~
\mathcal{D}^{tp,p}_\perp =
(\tilde{A}^{tp}_0+ \tilde{A}^{tp}_1) e^{-\mu^{tp,p}d} + 
\frac{a^{tp,p}_\perp}{d^4} ,\nonumber\\
&&\mathcal{D}^{pp,t}_\perp =
(A^{pp}_0+ A^{pp}_1) e^{-\mu^{pp,t}d} + \frac{a^{pp,t}_\perp}{d^4} ,\nonumber\\
&&\mathcal{D}^{pp,p}_\parallel =
\left[ A^{pp}_0+ A^{pp}_1 \left(1-\frac{1}{2}\mu^{pp,p}d\right) \right]
e^{-\mu^{pp,p}d} + \frac{a^{pp,p}_\parallel}{d^4} ,
\eeq
with the constraint $\tilde{A}^{tp}_0 + \tilde{A}^{tp}_1 = A^{tp}_0 + A^{tp}_1$,
meaning that, at $d=0$, the nonperturbative part of the correlation
functions $\mathcal{D}^{tp}$ have the same value.
The dependence of the various parameters on $B$ is understood and will be
discussed in the next section on the basis of the numerical results obtained
in Ref. \cite{DMMN2016} by lattice simulations of $N_f=2$ QCD
(at zero temperature).

\section{Numerical investigation and discussion on the gluon condensate}
\label{lattice}

The correlator
$\mathcal{D}_{\mu\nu,\xi}(d) \equiv \mathcal{D}_{\mu\nu,\mu\nu}(x = d\hat{\xi})$
has been discretized through the following lattice observable
\cite{DP1992,npb97}:
\beq
\mathcal{D}^L_{\mu \nu,\xi}(d) = \left\langle 
\Tr\left[\Omega^\dagger_{\mu\nu}(x) S(x,x+d\hat{\xi}) 
\Omega_{\mu\nu}(x+d\hat{\xi}) S^\dagger(x,x+d\hat{\xi})\right] \right\rangle ,
\label{lattice_correlator_form}
\eeq
where $\Omega_{\mu\nu}(x)$ stands for the traceless anti-Hermitian part of the 
corresponding plaquette, i.e.,
$\Omega_{\mu\nu} \equiv
\frac{1}{2} ( \Pi_{\mu\nu} - \Pi^\dagger_{\mu\nu} )
- \frac{1}{6} \Tr [ \Pi_{\mu\nu} - \Pi^\dagger_{\mu\nu} ] {\bf I}$.
Of course, 
$\mathcal{D}^L_{\mu \nu,\xi}(d) \rightarrow a^4 \mathcal{D}_{\mu \nu,\xi}(d)$
when the lattice spacing $a \to 0$.

We have considered $N_f = 2$ QCD discretized via 
unimproved rooted {\it staggered} fermions and the standard plaquette action
for the pure-gauge sector.
The background magnetic field $\vec{B} = B \hat{z}$ couples to the quark
electric charges ($q_u = 2|e|/3$ and $q_d = -|e|/3$, $|e|$ being the
elementary charge) and its introduction corresponds to additional
$U(1)$ phases entering the elementary parallel transports
in the discretized lattice version.
Periodicity constraints impose the following condition of quantization on $B$:
$|e| B = {6 \pi b}/{(a^2 L_x L_y)}$, $b \in \mathbb{Z}$.

Numerical simulations have been performed on a $24^4$ lattice by means of
the {\it Rational Hybrid Monte Carlo} algorithm
\cite{Gottlieb:1987mq,Kennedy:1998cu}
implemented on GPU cards, 
with statistics of $O(10^3)$ molecular-dynamics time units
for each $b$, with $b$ ranging from $0$ to $18$ (corresponding to
$0 \leq |e|B \leq 1.46$ GeV$^2$).
The bare parameters have been set to $\beta=5.55$ and $am =0.0125$, 
corresponding to a lattice spacing $a \simeq 0.125$ fm and to a
pseudo-Goldstone pion mass $m_\pi \simeq 480$ MeV. 

In order to remove ultraviolet fluctuations, 
following the previous studies of the gluon-field correlators
\cite{DP1992,npb97,plb97,DDM2003},
a {\it cooling} technique has been used 
which, acting as a diffusion process, smooths out short-distance fluctuations
without touching physics at larger distances: for a correlator 
at a given distance $d$, this shows up as an approximate 
plateau in the dependence of the correlator on the number of cooling steps,
whose location defines the value of the correlator.




For each value of $|e|B$, we have fitted the correlators with the
parametrization (\ref{newparam}), including distances in the range
$3\leq d/a \leq 8$, thus obtaining an estimate for all 
parameters.
From these best fits, it has emerged that the $8$ parameters pertaining to the
perturbative part of the correlation functions (\ref{newparam}) satisfy,
within the errors, the following equalities: 
\beq
a^{tt,t}_\parallel \simeq a^{tp,t}_\parallel \simeq a^{tp,p}_\parallel \simeq
a^{pp,p}_\parallel \equiv a_\parallel ,~~~~
a^{tt,p}_\perp \simeq a^{tp,t}_\perp \simeq a^{tp,p}_\perp \simeq
a^{pp,t}_\perp \equiv a_\perp ,
\label{pert_assumption}
\eeq
and, moreover, their dependence on $B$ is negligible.
In other words, the perturbative part of the correlators shows no
significant departure from the case $B=0$ [see Eq. (\ref{oldparam})]:
therefore, we have fitted again all the data with the parametrization
(\ref{newparam}) together with the assumption (\ref{pert_assumption}).

Concerning the parameters $\mu$ [i.e., the inverse of the
{\it correlation lengths} in the exponential terms in Eq. (\ref{newparam})],
they show a general tendency for a modest increase, which amounts
to about $5-10\%$ for the largest values of $|e|B$ and is slightly more
visible for the correlators in the directions orthogonal to $\vec{B}$.

\begin{figure}[t!]
\includegraphics[width=0.92\columnwidth, clip]{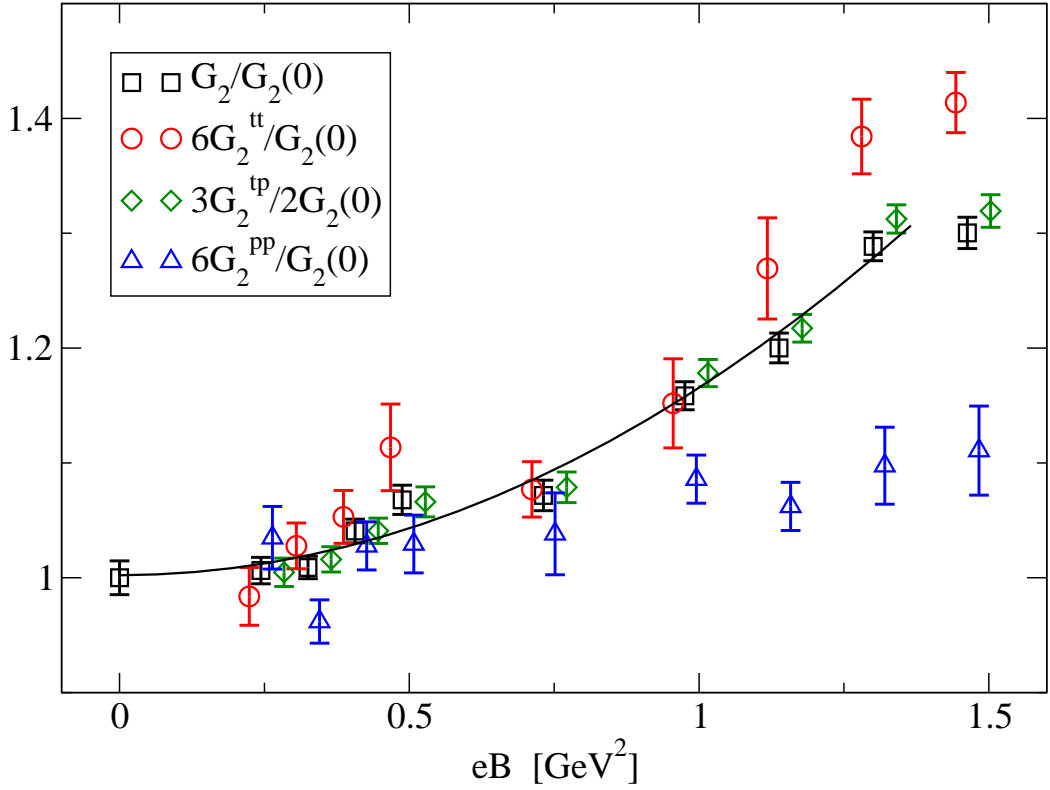}
\caption{Effects of the magnetic field on the gluon condensate $G_2$ and
on the three different contributions $G_2^{tt}$, $G_2^{tp}$ and $G_2^{pp}$.
The data points are shifted horizontally for the sake of readability.}
\label{g2}
\end{figure}

Among the various parameters entering Eq.~(\ref{newparam}), the ones
showing the most pronounced variation with $|e| B$ have been the 
nonperturbative coefficients $A_0$ and $A_1$.
That implies a significant dependence on the magnetic field 
of the {\it gluon condensate}, which is defined as:
\beq
G_2 = \frac{g^2}{4\pi^2} \sum_{\mu \nu, a}
\langle G^a_{\mu\nu} G^a_{\mu\nu} \rangle
\label{g2def}
\eeq
and is related to the correlator in Eq.~(\ref{defcorr}) through an
{\it Operator Product Expansion}. It encodes the main effect of
nonperturbative physics to gluon dynamics and its relevance was first
pointed out in Ref. \cite{SVZ79}.
One can extract $G_2$ from the small-distance limit of the nonperturbative part
of the correlator, obtaining, using our parametrization (\ref{newparam}):
\beq
G_2 = \frac{1}{\pi^2} \left[ \left( A^{tt}_0 + A^{tt}_1 \right)
+ 4 \left(A^{tp}_0 + A^{tp}_1 \right)
+ \left( A^{pp}_0 + A^{pp}_1 \right) \right]
\equiv G_2^{tt} + G_2^{tp} + G_2^{pp} ,
\label{gluonconda}
\eeq
where, in the last passage, we have distinguished three contributions, coming
from different sets of plaquettes.
In Fig. \ref{g2} we report the values obtained for $G_2$ as a function of
$|e|B$, normalized to the value of the condensate obtained for $B = 0$,
where we obtain {$G_2=3.56(5)\cdot 10^{-2}$ GeV$^4$},
the reported error being just the statistical one.

We notice that $G_2$ grows as a function of $|e|B$, the increase being
of the order of $25\%$ for the largest value of $|e|B$ explored.
In the same figure we also report the relative increases in the
$G_2^{tt}$, $G_2^{tp}$ and $G_2^{pp}$ terms. We see that the $tt$ term is the
most affected by the magnetic field, whereas the $pp$ contribution shows a
really modest dependence on $|e|B$.
In Fig. \ref{g2}, the best fit  with a quadratic 
function $G_2(|e|B)/G_2(0) = 1 + K (|e|B)^2$ is also plotted.
We obtain $K=0.164(7)\ \textnormal{GeV}^{-4}$ and
$\chi^2/n_{\rm d.o.f.} = 1.52$, excluding the point at $|e|B=1.46$ GeV$^2$.

An increase  of the chromomagnetic gluon condensate with $|e|B$ 
has been also found in Ref. \cite{ozaki}, which is  in qualitative agreement 
with the result presented here. A similar behaviour for $G_2$ has been also 
predicted making use of QCD sum rules \cite{Ayala:2015qwa}.

\section{Conclusions}
\label{finally}

We have found evidence of a significant effect of the external magnetic field
on the gluon-field correlation functions.
In particular, a large effect (and a significant anisotropy) is observed for
the coefficients of the nonperturbative terms in Eq. (\ref{newparam}),
which, on the basis of Eq. (\ref{gluonconda}), can be directly
related to the gluon condensate.
Due to the explicit Lorentz/$SO(4)$ symmetry breaking caused by the magnetic
field, we can distinguish among three different contributions to the gluon
condensate. An analysis based on Eq.~(\ref{gluonconda}) shows that each term has
a different behaviour as a function of the magnetic field (see Fig.~\ref{g2}).
Starting from that, we have observed that the gluon condensate itself increases
as a function of $B$, with the increase being of the order of 20\%
for $|e|B \sim 1$ GeV$^2$. Relative differences between the different
contributions are of the same order of magnitude, meaning that
anisotropies induced by $B$ are significant and comparable
to those observed in other pure-gauge quantities
(see, e.g., Ref.~\cite{struzzo}).
The increase of the gluon condensate provides evidence of the phenomenon
known as {\it gluon catalysis}, which had been previously observed
analysing the magnetic-field effects on plaquette expectation values
\cite{Ilgenfritz:2012fw,reg2,Ilgenfritz:2013ara}.




\end{document}